\documentclass[12pt]{article}
\usepackage{amsmath,amssymb,amsthm,graphicx,psfrag}
\usepackage{enumerate,subfigure}
\usepackage{pstricks,pst-node,multido}

\newtheorem{thm}{Theorem}[section]

\theoremstyle{definition}
   
   \newtheorem{ex}[thm]{Example}

\theoremstyle{remark}

\numberwithin{thm}{section}
\numberwithin{equation}{section}

\title{A number theoretical observation about the degeneracy of the genetic code}
\author{Andrei Khrennikov  and Marcus Nilsson \\ International Center for
Mathematical Modeling\\ in Physics and Cognitive Sciences, \\
University of V\"axj\"o, S-35195, Sweden\\
Email: Andrei.Khrennikov@msi.vxu.se}

\begin{document}
\maketitle

\begin{abstract} 
We discuss the similarity of the degeneration structure of the genetic code with 
a pure number theoretic -- ``divisors code.'' The most interesting thing about our observation is not that there is
a connection between number theory and the genetic code, but the
simlicity of the rule. We hope that the observation and the naive model presented in this
paper will serve for ideas to other models of the degeneracy of the genetic
code. Maybe, the ideas of this article can also be used in the area of
artificial life to syntesize artificial genetic codes.
\end{abstract}

\section{Introduction}

Recently there was shown a lot of interest to number theoretic studies 
of the problem of degeneration of the genetic code, 
see \cite{pitkanen06}-- \cite{KHRM}. In this paper we emphasize the similarity of 
the degeneration structure of the genetic code and a so called ``divisors code.''
The latter one arises very naturally from purely number theoretic arguments. 

\section{Genetic code}

The information contained in the DNA molecule is translated into
aminoacids by the messenger and the transcription RNA. The aminoacids
form protein molecules. A sequence of three nukleotides in the mRNA
molecule, a so called
codon, encodes for a specific aminoacid. However, there are different
codons that encode for the same aminoacid.  
This degeneracy of the genetic code is still not fully understood.
In this article we mean the so called standard genetic code when we
talk about genetic code. We have used Wikipedia \cite{wikipedia} to
get information about the genetic code.

In this article it will be announce an unpredicted similarity  between the
degeneracy of the genetic code and the number of positive divisors of
the first the first $20$ integers. The idea that there are some
similarities beteen the number theory and the genetic code is not
new. In \cite{pitkanen06} the set of the first $18$ prime numbers
together with $0$ och $1$ is used for representing the aminoacids.

We also present a model (based on the observation) that simulates an artificial genetic code.

\section{The number theoretic observation about the genetic code}

Let $A$ be the set of aminoacids. We let $A(k)$ denote the number of
aminoacids in $A$ that are encoded by $k$ codons. 
If we use the standard genetic code we get the following table.
\begin{center}
\begin{tabular}{|l|l|}
\hline
$k$ & $A(k)$\\
\hline
1  &   2 \\
2  &   9 \\
3  &   1 \\
4  &   5 \\
5  &   0 \\
6  &   3 \\
\hline
\end{tabular}
\end{center}
Now, let $B=\{1,2,\ldots, 20\}$. The number of positive divisors of a
positive integer $m$ is denoted by $\tau(m)$. By $B(k)$ we mean the
number of elements in $B$ that have $k$ positive divisors. Hence
\[
 B(k)=\{b\in B;\tau(b)=k\}.
\]
Let us add $B(k)$, $1\leq k\leq 6$ as a third column in the table above. 
\begin{center}
\begin{tabular}{|l|l|l|}
\hline
$k$ & $A(k)$ & $B(k)$ \\
\hline
1  &   2  &    1\\
2  &   9  &    8\\
3  &   1  &    2\\
4  &   5  &    5\\
5  &   0  &    1\\
6  &   3  &    3\\
\hline
\end{tabular}
\end{center}
The match between the second and the third column is not absolut but
it is still remarkable. We use this observation to construct an
elementary ``genetic code like model'', abbreviated GCLM, in the next section.

In the standard genetic code there are $3$ codons, that do not encode
for an aminoacid, they encode for that the end of the protein is
reached. We say that these codes encode for STOP.
If we add the STOP to the set of aminoacids and let
$B=\{1,2,\ldots,21\}$ we get the following table:
\begin{center}
\begin{tabular}{|l|l|l|}
\hline
$k$ & $A(k)$ & $B(k)$ \\
\hline
1  &   2  &    1\\
2  &   9  &    8\\
3  &   2  &    2\\
4  &   5  &    6\\
5  &   0  &    1\\
6  &   3  &    3\\
\hline
\end{tabular}
\end{center}

\section{A number theoretical ``genetic code''}

The number of positive divisors of a positive integer $m$ is the same
as the number of pairs of the form $(a,b)$, where $a,b\geq 1$ and
$a\cdot b=m$. This can be visualized as the number of rectangles with
integer side lengths and area $m$ that can be formed.

The idea is that the numbers $1,2,\ldots, 20$ represent $20$ different
aminoacids. The pairs of positive numbers  of the form
$(a,b)$, where $ab=m$, represents the codons that code for the
aminoacid represented by $m$. We call such a code {\it divisors code.}

Of course, it is easy to generalize this
to any number of aminoacids (building blocks).

\begin{ex}
The code sequence
\begin{equation}
 ((1,3),(5,2), (3,1), (2,1), (2,3), (1,4), (11,1), (6,3))
\end{equation}
encode for the sequence
\[
 (3,10,3,2,6,4,11,18).
\]  
A possible visualization of the code sequence is shown in 
Figure~\ref{rek_code1}.
\begin{figure}
\begin{center}
  \psset{unit=0.3cm}
  \begin{pspicture}(0,0)(25,11)
    \psframe*(0,0)(3,1)
    \psframe*(4,0)(6,5)
    \psframe*(7,0)(8,3)
    \psframe*(9,0)(10,2)
    \psframe*(11,0)(14,2)
    \psframe*(15,0)(19,1)
    \psframe*(20,0)(21,11)
    \psframe*(22,0)(25,6)
  \end{pspicture}
\end{center}
\label{rek_code1}
\caption{}
\end{figure}
\end{ex}

\begin{ex}
Let us consider a toy genetic model in that the alphabet has just two letters, say
$A=0$ and $B=1,$ codons have the length 3: $AAA, AAB, ..., BBB.$  There are totally 
8 codons. We suppose that they should encode 4 aminoacids: $B=\{1,2,3,4 \}.$
Aminoacids are encoded as follows: $1=(1,1), 2=\{(1,2), (2,1)\}, 3=\{ (3,1), (1,3\},
4=\{(1,4), (4,1), (2,2) \}.$ Here (in contrast to the standard genetic code)  
the total number of codons matches with the total number 
of divisors.
\end{ex}

\section{Discussion}
The most interesting thing about our observation is not that there is
a connection between number theory and the genetic code, but the
simlicity of the rule.

We hope that the observation and the naive model presented in this
paper will serve for ideas to other models of the degeneracy of the genetic
code. Maybe, the ideas of this article can also be used in the area of
artificial life to syntesize artificial genetic codes.

The divisors code which was presented in this article  would match precisely with the 
genetic code if for any $k$ the number $A(k)$ were equal to the number $B(k).$ We see
that there are deviations. Hence, Nature did not follow precisely to the laws of 
number theory. There might happen some mutations which disturbed the genetic code.

Results of this note were announced in \cite{KHRM1}

\end{document}